\documentclass[twocolumn,english,showpacs]{revtex4}
\usepackage{babel,amsmath,amssymb,dcolumn}
\usepackage[dvips]{graphics}

\def\agt{\mathrel{\raise.3ex\hbox{$>$}\mkern-14mu\lower0.6ex\hbox{$\sim$}}}
\def\alt{\mathrel{\raise.3ex\hbox{$<$}\mkern-14mu\lower0.6ex\hbox{$\sim$}}}
\def\eps{\epsilon}
\def\beq{\begin{equation}}
\def\eeq{\end{equation}}
\def\bsubeq{\begin{subequations}}
\def\esubeq{\end{subequations}}

\def\pd{{\partial}}

\begin{document}

\title{Perturbations of Schwarzschild black holes in laboratories}

\author{E. Abdalla}
\email{eabdalla@fma.if.usp.br}
\author{R.A. Konoplya}
\email{konoplya@fma.if.usp.br}
\author{A. Zhidenko}
\email{zhidenko@fma.if.usp.br}
\affiliation{Instituto de F\'{\i}sica, Universidade de S\~{a}o Paulo \\
C.P. 66318, 05315-970, S\~{a}o Paulo-SP, Brazil}


\begin{abstract}
It is well-known that the perturbations of Schwarzschild black
holes are governed by a wave equation with some effective
potential. We consider perturbations of a gas in a tube called the
de Laval nozzle, which is narrow in the middle and has a sonic
point in the throat. By equating the wave equation in a de Laval
nozzle of an arbitrary form with the wave equation of spin-s
perturbations of Schwarzschild black holes, we find the exact
expression for the form of the de Laval nozzle, for which acoustic
perturbations of the gas flow corresponds to the general form of
perturbations of Schwarzschild black holes. This allows
observation, in a laboratory, of the acoustic waves, which are
analogue of damping quasinormal oscillations of Schwarzschild
black holes. The found exact acoustic analog allows to observe
also some other phenomena governed by the wave equation, such as
the wave scattering and tunneling.
\end{abstract}

\maketitle

%%%%%%%%%%%%%%%%%%%%%%%%%%%%%%%%%%%%%%%%%%%%%%%%%%%%%%%%%%%%%%%%%%%%%%%%%%%%%%%
\section{Introduction}
%%%%%%%%%%%%%%%%%%%%%%%%%%%%%%%%%%%%%%%%%%%%%%%%%%%%%%%%%%%%%%%%%%%%%%%%%%%%%%%

One of the most essential characteristics of black holes are the
resonant frequencies of the response to external perturbations,
called quasinormal frequencies \cite{Kokkotas-99}. Being an analog
of normal modes for open systems, they do not depend on the way in
which the system is perturbed, but only on the parameters of a
system. Quasinormal modes are expected to be observed in the near
future with the help of a new generation of gravitational
antennas. It is well-known that non-rotating astrophysical black
hole can be described by the Schwarzschild solution, implying the
importance of the quasinormal modes of this background
\cite{SQNMs}.

In addition to the still elusive possibility to observe
quasinormal modes (QNMs) of black holes with the help of a new
generation of gravitational antennas, there is a window for
observation of the acoustic analogue of a black hole in
laboratories. This is the well-known Unruh analogue of black holes
\cite{Unruh}, which are the apparent horizons appearing in a
fluid with a space-dependent velocity, in the presence of sonic
points. The supersonic waves cannot propagate back beyond the
sonic point, mimicking thereby, the effect of the horizon at sonic
points in a membrane paradigm. The Unruh discovery stimulated
active investigation of quasinormal modes of different analogue
black holes \cite{analogueQNMs}. First of all, the quasinormal
spectrum of the analogue black holes given by the metrics:
$$ d s^{2} = -(1 - C/r^{2}) d t^{2} + (1- C/r^{2})^{-1} d r^{2} - 2 B d \phi d t + r^{2} d \phi^{2}$$
and
$$ d s^{2} = - (1- r_{0}/r^{4}) d t^{2} + (1- r_{0}/r^{4})^{-1} d r^{2} + r^{2} d \Omega^{2}$$
were considered  \cite{analogueQNMs}. These are the two models for
rotating analogue black holes in "draining bathtub" and for a
canonical analogue non-rotating black holes. Recently, interesting
acoustic analogues of brane-world black holes were suggested
\cite{Ge:2007tr,Ge:2007ts}.

We can see from the above formulas that these metrics, even being
very useful as analogues with apparent horizons, do not represent
a true solutions of the Einstein or other gravity dynamical
equations. If one had a complete analogy with some known solution
of Einstein equations, say, the Schwarzschild solution, he could
see in the acoustic experiments not only qualitative, but also, up
to an experimental accuracy, exact numerical coincidence with a
prototype characteristics. Namely, for quasinormal modes, which
are governed by the form of the wave equation, this numerical
correspondence would mean that the effective potential of the
perturbations of some hydrodynamic system coincides with an
effective potential of a black hole. Fortunately, recent
consideration of the perturbations of a gas in de Laval nozzle
\cite{Sakagami} gives us such an opportunity of finding a system that is realted to
the same effective potential as a Schwarzschild black hole.

The canonical de Laval nozzle is a convergent-divergent tube,
narrow in the middle. It allows to accelerate the gas until the
sonic speed in its throat, reaching supersonic speeds after
passing the throat. The perturbations of the gas in de Laval
nozzle can be considered as one-dimensional if the section of the
nozzle does not change too quickly. Here we show that the
corresponding effective potential of perturbations in a canonical
de Laval nozzle can be made to be equal to the potential for
perturbations of Schwarzschild black holes, if choosing some
specific form of the nozzle. In addition, we suggest another,
approximate, way to get quasinormal modes of Schwarzschild black
holes in de Laval nozzle. For this, one needs to mimic the form of
the effective potential for Schwarzschild metric with the help of
a de Laval nozzle of a simple form suggested in
\cite{Sakagami}.

The paper is organized as follows: Sec \ref{sec:basic} gives all
basic equations of the one-dimensional motion in de Laval nozzle
we shall use. Sec \ref{sec:mimic} is devoted to reproducing the
exact expression for the form of the nozzle which corresponds to
the potential of the Schwarzschild black holes.
%In Sec IV we propose alternative
%method for obtaining the Schwarzschild quaisnormal modes with a
%simple form of de Laval nozzle.
In the discussions, we sketch the open questions and possible
generalizations of the suggested technique.

\section{Calculation of the nozzle cross section $A$ in terms of $g$}\label{sec:basic}

We assume that a gas in the nozzle can be described by equations
of motion for perfect fluid and that the flow is
quasi-one-dimensional:
\begin{gather}
\pd_t(\rho A) + \pd_x(\rho vA) = 0 \,,
\label{eq:cont} \\
\pd_t(\rho vA) + \pd_x[(\rho v^2 + p)A] = 0 \,,
\label{eq:momentum} \\
\pd_t(\eps A) +  \pd_x[(\eps+p)vA] = 0 \,,
\label{eq:energy}
\end{gather}
Here $\rho$ is the density, $v$ is the fluid velocity, $p$ is the
pressure, $A$ is the cross section of the nozzle, and
\beq
\eps = \frac{1}{2}\rho v^2 + \frac{p}{\gamma-1}
\eeq
is the energy density. The heat capacity ratio is
$\gamma=1+2/n=7/5=1.4$ for di-atomic molecules of air ($n=5$). We
shall assume that the flow has no entropy discontinuity, then the
fluid is isentropic
\beq
p\propto\rho^\gamma \,.
\label{eq:isentropic}
\eeq
Instead of Eq.~\eqref{eq:momentum}, we can use Euler's equation
\begin{gather}
\rho(\pd_t + v \pd_x)v = -\pd_x p \,,
\label{eq:Euler}
\end{gather}
For isentropic fluid Eq.~\eqref{eq:Euler} is reduced to the
Bernoulli's equation
\beq
\pd_t\Phi + \frac{1}{2}(\pd_x\Phi)^2 + h(\rho) = 0 \,,
\label{eq:Bernoulli}
\eeq
where $h(\rho) \equiv \int\rho^{-1}dp$ is the specific enthalpy
and $\Phi = \int v\,dx$ is the velocity potential.

According to \cite{Sakagami}, the perturbation equations in such a
nozzle can be reduced to:
\begin{gather}
\biggl[ \frac{d^2}{dx^{*2}} + \kappa^2 - V(x^*) \biggr] H_\omega = 0, \label{eq:Sch1}\\
\kappa = \frac{\omega}{c_{s0}}, \\
V(x^*) = \frac{1}{g^2}\biggl[\; \frac{g}{2}\frac{d^2g}{dx^{*2}}
    - \frac{1}{4}\Bigl(\frac{dg}{dx^*}\Bigr)^2 \;\biggr].
\end{gather}
Here $c_{s0}$ is the stagnation sound speed, and $x^{*}$ is an
acoustic analogue of the tortoise coordinate which satisfies
$x^{*}(x=+\infty) = + \infty$, $x^{*}(x=0) = - \infty$, namely,
\begin{equation}
x^{*} = c_{s0} \int \frac{d x}{c_{s} (1-M(x)^{2})},
\end{equation}
where $M(x)$ is the Mach number \cite{Hydro}, which, by the
definition,  is the current flow speed divided by the sound speed.
In our notations $M = v/c_s$. The function $H_{\omega}$ represents
small perturbations of gas flow,
\begin{gather}
H_\omega(x) = g^{1/2}\int dt~e^{i\omega[t-f(x)]}\phi(t,x),
\end{gather}
\begin{gather}
g = \frac{\sigma}{c_s}, \\
f(x) = \int\frac{|v|\,dx}{c_s^2-v^2},
\end{gather}
Here, according to \cite{Sakagami}, the small perturbations are
defined as follows
\begin{align}
\rho &= \bar\rho + \delta\rho \,, \qquad \bar\rho \gg |\delta\rho|   \,,
\label{eq:split_rho} \\
\Phi &= \bar\Phi + \phi \,,       \qquad |\pd_x\bar\Phi| \gg |\pd_x\phi| \,,
\label{eq:split_Phi}
\end{align}

Our starting point is the calculation of  the configuration of de
Laval nozzle, i. e. its cross section as a function of the
transversal nozzle coordinate. Since we know the effective
potential, we can calculate in some way the function $g(x)$. By
definition Eq. (15) of \cite{Sakagami}
$$g=\frac{\sigma}{c_s}=\frac{\rho A}{\sqrt{\gamma p/\rho}}.$$
Taking (5) into account we find
\begin{equation}\label{gdef}
g\propto\frac{\rho A}{\rho^{(\gamma-1)/2}}.
\end{equation}

We can choose dimensionless quantities for $\rho(x)$ and $A(x)$ by
measuring them in units of $\rho_0$ and $A^*$ respectively
\cite{Hydro}. Then equation (5.3) of \cite{Hydro} reads
\begin{equation}\label{Adef}
A^{-1}\propto\left(1-\rho^{(\gamma-1)}\right)^{1/2}\rho.
\end{equation}

Since (\ref{eq:Sch1}) is invariant with respect to re-scaling of
$g$, we can fix the coefficients in (\ref{gdef}) and (\ref{Adef})
arbitrarily:
\begin{equation}\label{constfix}
g=\frac{\rho A}{2\rho^{(\gamma-1)/2}}, \quad
A^{-1}=\left(1-\rho^{(\gamma-1)}\right)^{1/2}\rho.
\end{equation}
We find
\begin{equation}
g=\frac{\rho^{(1-\gamma)/2}}{2\left(1-\rho^{(\gamma-1)}\right)^{1/2}}
=\frac{\rho^{(1-\gamma)}}{2\left(\rho^{(1-\gamma)}-1\right)^{1/2}}
\end{equation}
Hence it follows that
\begin{equation}
\rho^{1-\gamma}=2g^2\left(1\pm
\sqrt{1-g^{-2}}\right).
\end{equation}
The sign should be chosen in order that $\rho$ be a monotonous
function with respect to the transverse coordinate. As we will
show later, the function $g$ for the Schwarzschild black hole can
be chosen also monotonous in the $R$ region, finite at the horizon
and infinite at the spatial infinity. Therefore, we choose the
minus sign,
\begin{equation}
\rho^{1-\gamma}=2g^2\left(1-\sqrt{1-g^{-2}}\right).
\end{equation}
Note that $g$ must always be larger than unity in our
consideration.

In our notations, the Mach number is connected with $\rho$ as
\begin{equation}\label{rhosolution}
\rho^{1-\gamma}=1+\frac{\gamma-1}{2}M^2.
\end{equation}
We find
$$M^2=\frac{2}{\gamma-1}\left(\rho^{1-\gamma}-1\right)=$$
\begin{equation}
\frac{2}{\gamma-1}\left(2g^2\left(1-\sqrt{1-g^{-2}}\right)-1\right).
\end{equation}
Since $M=1$ at the event horizon, $g$ must be finite there, and
\begin{equation}\label{normal}
g\Biggr|_{e.h.}=\frac{\gamma+1}{2\sqrt{2}\sqrt{\gamma-1}}=\frac{3}{\sqrt{5}}>1.
\end{equation}
This requirement fixes both constants of integration.

Substituting (\ref{rhosolution}) in (\ref{constfix}) we find the
cross-section area as a function of $g$:
\begin{equation}\label{2}
A=\frac{\sqrt{2}\left(2g^2\left(1-\sqrt{1-g^{-2}}\right)\right)^{1/(\gamma-1)}}{\sqrt{1-\sqrt{1-g^{-2}}}}.
\end{equation}

\section{The de Laval nozzle for the Schwarzschild black hole}\label{sec:mimic}

Our starting point is the calculation of the configuration of the
de Laval nozzle, i. e. its cross section as a function of the
transversal nozzle coordinate. Since we know the effective
potential, we can calculate in some way the function g(x).

\begin{figure*}\label{1figure}
\caption{The form of de Laval nozzle and the effective potential for $s=\ell=0$.}
\resizebox{\linewidth}{!}{\includegraphics*{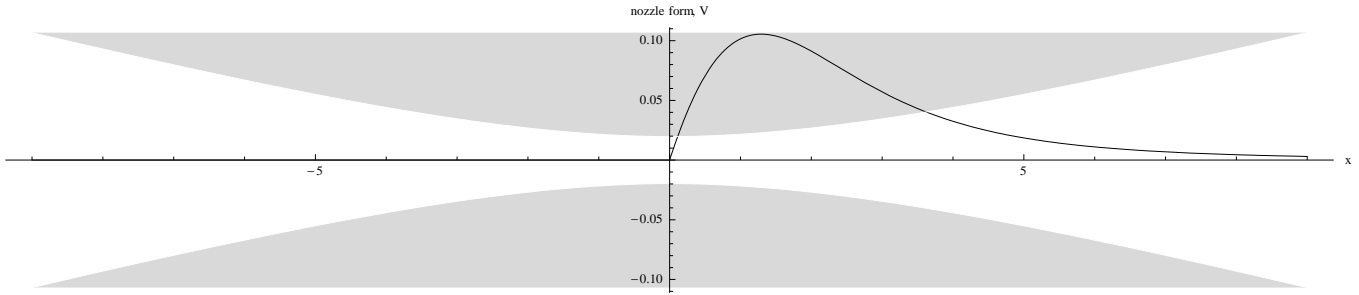}}
\end{figure*}

\begin{figure*}\label{2figure}
\caption{The form of de Laval nozzle and the effective potential for $s=\ell=1$.}
\resizebox{\linewidth}{!}{\includegraphics*{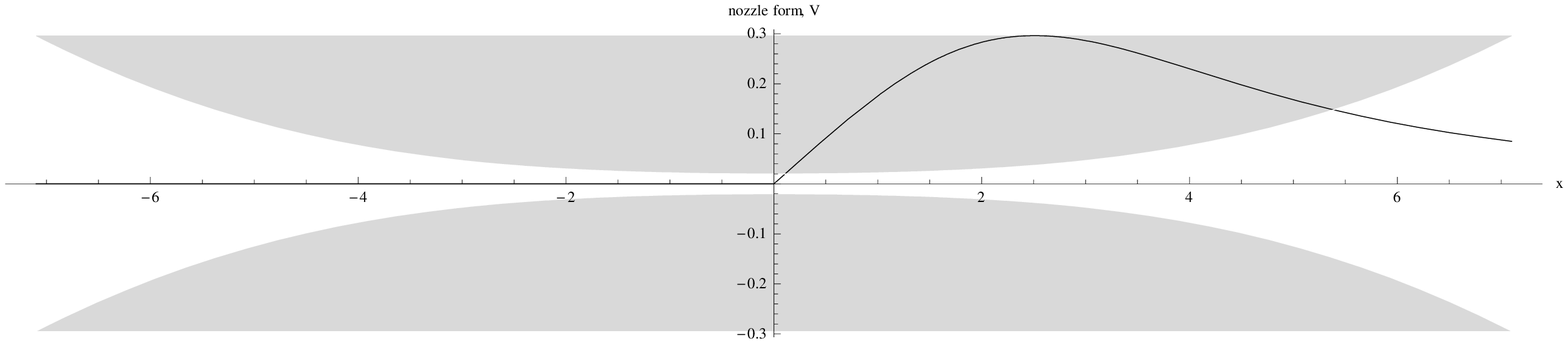}}
\end{figure*}

\begin{figure*}\label{3figure}
\caption{The form of de Laval nozzle and the effective potential for $s=\ell=2$.}
\resizebox{\linewidth}{!}{\includegraphics*{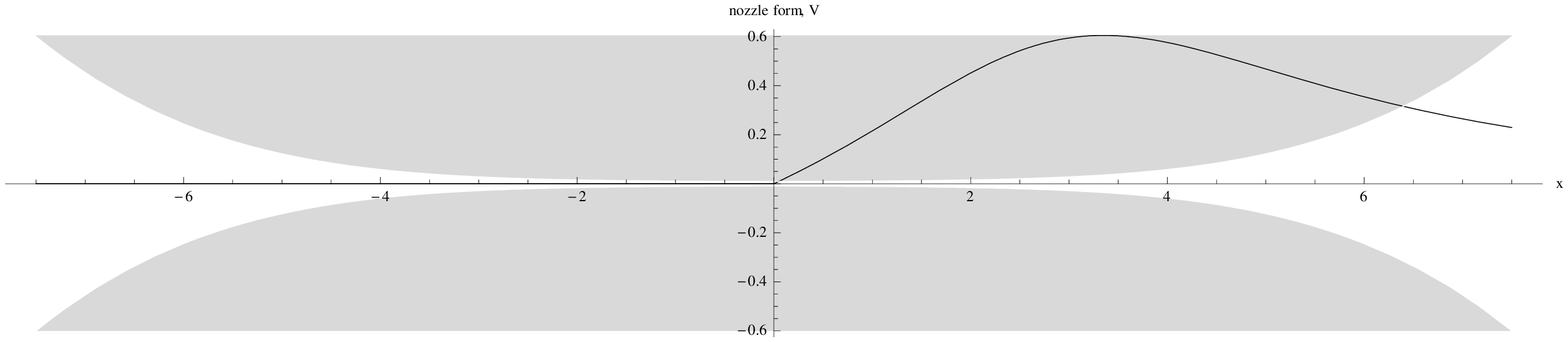}}
\end{figure*}

After separation of the angular and time variables, scalar field
perturbations in the Schwarzschild background, can be reduced to
the wave-like equation
\begin{equation}
\left(\frac{d^2}{dr_*^2}+\omega^2-V(r^*)\right)\Psi(r^*)=0,
\end{equation}
where putting the event horizon to be unity we find,
$$f(r)=1-\frac{1}{r}, \quad dr^*=\frac{dr}{f(r)}$$
\begin{equation}
V(r)=f(r)\left(\frac{\ell(\ell+1)}{r^2}+\frac{1-s^2}{r^3}\right)
\end{equation}

To find $g$ that produces the same potential we identify the
"tortoise" coordinates of the black hole solution and of the laval
nozzle
$$dr^*=dx^*=\frac{c_{s0}dx}{c_s(1-M^2)}=\frac{\rho^{(1-\gamma)/2}dx}{1-M^2}=$$
\begin{equation}\label{identify}
\frac{\sqrt{2g^2\left(1-\sqrt{1-g^{-2}}\right)}dx}{1-\frac{2}{\gamma-1}
\left(2g^2\left(1-\sqrt{1-g^{-2}}\right)-1\right)}.
\end{equation}
Here $x$ is the real coordinate along de Laval nozzle. Then we can
find the equation for $g(r)$,
\begin{equation}\label{gequation}
\frac{f(r)f'(r)g'(r)+f(r)^2g''(r)}{2g(r)}-\frac{f(r)^2g'(r)^2}{4g(r)^2}=V(r).
\end{equation}
This implies that the from of de Laval nozzle is parameterized by
the parameter $r$ and  $x^{*}(r)$ = $r^{*}(r)$. Note that as we
chose the radius of the event horizon to be unity, the nozzle
coordinate $x$ is measured in the units of the radius of the event
horizon.

%The symmetry of the equation $\delta g(r)= C g(r)$ allows
%decreasing of the equation order
%$$g(r)=\exp\left(2\int W(r^*)dr^*\right)=\exp\left(2\int \frac{W(r)}{f(r)}dr\right),$$
%\begin{equation}\label{order}
%V(r)=\frac{dW}{dr^*}+W(r^*)^2=f(r)W'(r) + W(r)^2.
%\end{equation}

The general solution of the equation (\ref{gequation}) contains
two arbitrary constants. They can be fixed in a unique way by the
condition (\ref{normal}). Namely, the requirement that the
solution must be finite at $r=1$  fixes one of the constant. Then
the other constant re-scales the solution of (\ref{gequation}),
and must be fixed by its value at $r=1$. Finally, the solution of
(\ref{gequation}) for arbitrary $\ell$ and $s$, that satisfies
(\ref{normal}) is given by the following formula:
$$ g(r) = $$
\begin{eqnarray}\nonumber\label{1}
&&\frac{\gamma+1}{2\sqrt{2}\sqrt{\gamma-1}}\sum_{n=s}^{\ell}
\left(\frac{(-1)^{n+s} (\ell + n)!}{( n + s )! ( n - s )! (\ell - n)!}r^{n+1}\right)^2 = \\
&&=\small\frac{\gamma+1}{2\sqrt{2}\sqrt{\gamma-1}}r^{2s+2}\times
\\\nonumber&&\times\left(\frac{\Gamma(1+\ell+s)_2F_1(s-\ell,s+\ell+1,1+2s,r)}
{\Gamma(1+\ell-s)\Gamma(1+2s)}\right)^2.
\end{eqnarray}
One can easily check that the above solution indeed satisfies the
equation (\ref{normal}), for any fixed $\ell$ and $s$.

From (\ref{identify}) we find the dependance of the transversal
nozzle coordinate $x$ on the parameter $r$:
\begin{equation}
x=\intop_1^r\frac{\left(\gamma+1-4g(r)^2\left(1-\sqrt{1-g(r)^{-2}}\right)\right)dr}
{f(r)(\gamma-1)\sqrt{2g(r)^2\left(1-\sqrt{1-g(r)^{-2}}\right)}}.
\end{equation}
The integration constant is chosen in order to $x$ be zero at the sonic point.

Now we are in position to find the required form of de Laval
nozzle. i.e. to find its cross-section $A(x)$. We just need to
replace $g(r)$ given in (\ref{1}) in (\ref{2}) and go over to the
transverse nozzle coordinate $x$. The function $A(x)$ is shown in
Figs. 1 - 4. Note that the canonical de Laval nozzle is diverging
at the end of the flow trajectory, so that $A_{x=\infty} =\infty$
(see pages $53$ and $124$ in \cite{Hydro}). Indeed, our formula
(\ref{1}) implies divergence at least as $\sim r^2$. The diverging
of the nozzle nevertheless does not give any going beyond the
one-dimensional representation of the motion, because the function
$\sqrt{A(x)}$ is measured in units of black hole mass, i.e. one
can "pull" the nozzle along the transverse coordinate $x$ in order
to make the area of the nozzle change as slowly as one wishes.
Such a "pulling" simply means that we are getting the
correspondence with a black hole of larger mass. Since the
quasinormal modes are inversely proportional to the mass of the
black hole, this means just some determined multiplication by a
coefficient when coming from the frequencies observed in
experiment to the QNMs of a black hole.

As can be seen from \cite{KZHPLB2}, the values of the quasinormal
modes are determined by the behavior of the effective potential in
some region near black hole. The form of the effective potential
(and thereby of de Laval nozzle) far from black hole is less
significant for the QNMs problem. Therefore we expect that such
experimental phenomena as surface friction and reflection of waves
from boundaries will not have considerable influence on the
observed picture.

\begin{figure*}\label{figure4}
\caption{The form of de Laval nozzle and the effective potential for the polar gravitational perturbation $\ell=2$.}
\resizebox{\linewidth}{!}{\includegraphics*{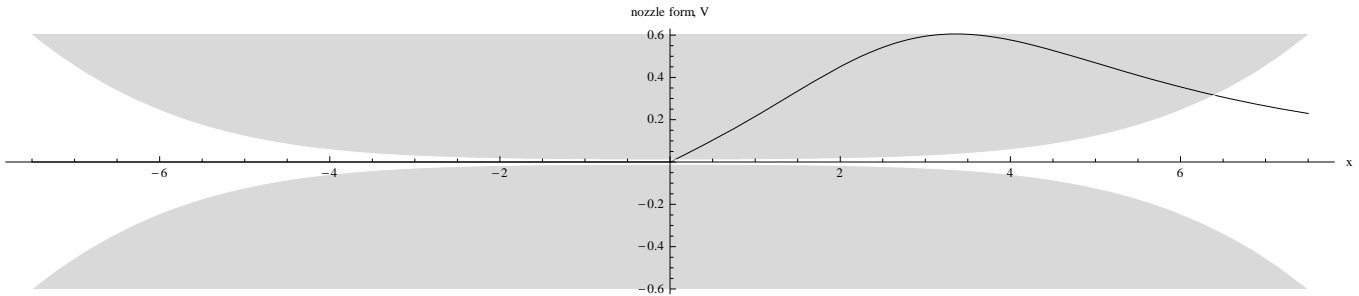}}
\end{figure*}

%isospectrality of two effective potentials, given by
%\begin{equation}\label{isopotential}
%V(r^*)=W(r^*)^2\pm\frac{dW}{dr^*}+\beta,
%\end{equation}
%that takes place e. g. for the potentials of axial and polar gravitational perturbations.

Now we discuss the isospectrality and the effective potential for
the polar and axial gravitational perturbations. We consider the
effective potential for the gravitational perturbation of the
polar type,
\begin{equation}\label{polar}
V(r)=f(r)\frac{9(1+\lambda r)+\lambda^3r^3+\lambda^2r^2(3+2r)}{r^3(3+\lambda)^2},
\end{equation}
where $\lambda = (\ell+2)(\ell-1)$.

For the above polar type gravitational perturbations we can also
obtain the exact solution for the function $g(r)$, i.e. the form
of de Laval nozzle. The function $g(r)$ is given in the following
table for $\ell =2$, $3$, $4$, and $5$ and in the formula
(\ref{g}):

\begin{equation}\label{g}
g(r,\ell)=\frac{\gamma+1}{2\sqrt{2}\sqrt{\gamma-1}}\frac{r^2p(r,\ell)^2}{(3+(\ell+2)(\ell-1)r)^2}
\end{equation}

\begin{tabular}{|c|l|}
\hline
$\ell$&$p(r,\ell)$\\
\hline
$2$&$3-6r^2-4r^3$\\
$3$&$3-30r^2-20r^3+60r^4$\\
$4$&$3-90r^2-60r^3+630r^4-504r^5$\\
$5$&$3-210r^2-140r^3+3570r^4-6552r^5+3360r^6$\\
%$6$&$3-420r^2-280r^3+14490r^4-44856r^5+50820r^6-19800r^7$\\
\hline
\end{tabular}

\vspace{3mm}

Apparently there exists a solution for general $\ell$, yet
probably quite cumbersome. Analysis of the Figs. 3-4 shows that
the form of the nozzles for modeling polar and axial gravitational
perturbations are almost the same. The difference cannot be seen
explicitly, although not vanishing. As the effective potentials
for axial and polar types also differ only slightly, this may mean
that the forms of the nozzles also differ only slightly.

\section{Discussion}

The suggested solution of the inverse problem for the
correspondence of the form of de Laval nozzle to the general form
of perturbations of the Schwarzschild black holes (i.e. for
perturbations with spin $s$ and multipole $\ell$) can be
generalized in many ways. First of all, it would be very
interesting to consider the massive vector \cite{vector} and
scalar \cite{scalar} field perturbations, because of quite unusual
behavior of massive perturbations. Thus the so-called
quasi-resonances, infinitely long lived modes, for massive scalar
field \cite{quasiresonances} could be observed in a de Laval
nozzle of some form as almost non-damping sound waves. These waves
would certainly be reflected from the boundary of the nozzle and,
thereby, would break the quasinormal mode boundary conditions.
Even though in a real experiment one cannot obtain perfect QNM
boundary condition (QNM b.c.), a considerable deviation from QNM
b.c. should be observed when modeling the quasi-resonances.

Another possible generalization is to consider more general black
hole backgrounds: Reissner-Nordstr\"om, Schwarzschild-de Sitter,
or higher dimensional Schwarzschild black holes \cite{higherD}
with charge, $\Lambda$-term and Gauss-Bonnet-term \cite{GB},
including the brane-world black holes
\cite{braneQNMs}. The Gauss-Bonnet black hole
ringing \cite{GB} would be especially interesting to model in a
nozzle because there exists the instability in some region of
values of the black hole parameters \cite{GB}. In the approach
considered in this paper we are limited only by a spherical
symmetry, i.e. by $\omega$ independence of the effective
potential. In addition, one could consider the flow of gas with a
time dependent initial speed at the compressor, which probably
could model the perturbations of the Vaidya evaporating black
holes \cite{Abdalla:2007hg}, \cite{Abdalla:2006vb}. We cannot be
sure, that in all these cases the differential equation for the
form of de Laval nozzle will be exactly integrable, yet one can
always find some numerical solution. We believe that further
research will solve these interesting problems.

It should be recalled also that the \emph{precise} acoustic
analogy is only established for a scalar field. To be able to
reproduce the potential $V(r)$ for fields of different spins
certainly does not mean that one can reproduce all the
characteristics of those equations in an acoustic model.

Finally, let us note that the obtained acoustic analogue for the
perturbations of the Schwarzschild black holes is not limited by
quasinormal mode problems only, but allow general investigation of
propagation of classical and quantum fields, including such
processes as scattering and tunnelling of waves and particles.

\begin{acknowledgments}
This work was supported by \emph{Funda\c{c}\~{a}o de Amparo
\`{a} Pesquisa do Estado de S\~{a}o Paulo (FAPESP)} and
\emph{Conselho Nacional de Desenvolvimento Cient\'ifico e Tecnol\'ogico (CNPq)},
Brazil.
\end{acknowledgments}

%%%%%%%%%%%%%%%%%%%%%%%%%%%%%%%%%%%%%%%%%%%%%%%%%%%%%%%%%%%%%%%%%%%%%%%%%%%%%%%

\end{document}